\documentclass[aps,prb,twocolumn,floatfix]{revtex4-1}
\usepackage[latin1]{inputenc}
\usepackage[english]{babel}
\usepackage{graphicx}
\usepackage{multirow}
\usepackage{float}
\usepackage{units}
\usepackage{subfigure}
\usepackage[cm]{fullpage}
\usepackage{dcolumn}
\usepackage{amsmath}
\begin{document}
\title{Scattering of flexural acoustic phonons at grain boundaries in graphene}
\author{Edit E. Helgee}
\affiliation{Department of Applied Physics, Chalmers University of Technology, SE-412 96, G\"oteborg, Sweden}
\author{Andreas Isacsson}
\affiliation{Department of Applied Physics, Chalmers University of Technology, SE-412 96, G\"oteborg, Sweden}

\begin{abstract}
We investigate the scattering of long-wavelength flexural phonons against grain boundaries in graphene using molecular dynamics simulations. Three symmetric tilt grain boundaires are considered: one with a misorientation angle of $17.9^\circ$ displaying an out-of-plane buckling \unit[1.5]{nm} high and \unit[5]{nm} wide, one with a misorientation angle of $9.4^\circ$ and an out-of-plane buckling \unit[0.6]{nm} high and \unit[1.7]{nm} wide, and one with a misorientation angle of $32.2^\circ$ and no out-of-plane buckling. At the flat grain boundary, the phonon transmission exceeds \unit[95]{\%} for wavelengths above \unit[1]{nm}. The buckled boundaries have a substantially lower transmission in this wavelength range, with a minimum transmission of \unit[20]{\%} for the $17.9^\circ$ boundary and \unit[40]{\%} for the $9.4^\circ$ boundary. At the buckled boundaries, coupling between flexural and longitudinal phonon modes is also observed. The results indicate that scattering of long-wavelength flexural phonons at grain boundaries in graphene is mainly due to out-of-plane buckling. A continuum mechanical model of the scattering process has been developed, providing a deeper understanding of the scattering process as well as a way to calculate the effect of a grain boundary on long-wavelength flexural phonons based on the buckling size.
\end{abstract}

\maketitle

\section{Introduction}
Due to its two-dimensional nature and exceptional mechanical properties, graphene is considered an interesting material for phononics\cite{Balandin2012phononics}. However, the use of graphene in phononics requires methods for engineering the vibrational properties. One approach might be to use extended defects, such as the grain boundaries known to exist in graphene grown by chemical vapor deposition\cite{Huang2011TEM}. These grain boundaries have been found to consist mainly of pentagon and heptagon defects and to have a tendency towards out-of-plane buckling\cite{Coraux2008buckle,Carlsson2011gbstruct,Liu2011gbstruct}. The electronic \cite{Yazyev2012energy,Zhang2013nonsym} and mechanical \cite{Lee2013strength, Cao2012mechanical} properties of grain boundaries have been investigated previously, but the effects on vibrational properties and phonon transport are still not known.

Diffusion of phonons is the dominating mechanism of thermal transport in graphene, and it has been claimed that in suspended graphene flexural acoustic phonons are particularly important\cite{Seol2010scienceZA,Lindsay2010thermal}. Phonon scattering at grain boundaries will therefore affect the thermal conductivity. The thermal conductivity across grain boundaries has previously been studied using non-equilibrium molecular dynamics and Green's function methods. It was found that grain boundaries in graphene reduce the thermal conductivity, although the effect is small compared to other materials \cite{Cao2012Kapitza,Cao2012asymmetric,Bagri2011,Lu2012Greens,Serov2013greens}. However, these studies do not provide detailed insight into the phonon scattering mechanism. 

In the present study, molecular dynamics is used to investigate the scattering of flexural (out-of-plane) phonons against grain boundaries in graphene. Three grain boundaries have been considered: one that shows no buckling, one that displays buckling with a height of \unit[0.6]{nm} and a width of \unit[1.7]{nm}, and one with a buckling height of \unit[1.5]{nm} and a buckling width of \unit[5]{nm}. We find that the grain boundary with the higher buckling may transmit as little as \unit[20]{\%} of the incoming phonon pulse at wavelengths above \unit[1]{nm}. For the boundary with the lower buckling the minimum transmission is \unit[40]{\%}, while the transmission at the flat boundary approaches 100 \% for the same wavelengths. Also, we find that the incoming flexural vibrations give rise to a longitudinal vibration when interacting with the buckled grain boundaries.

A continuum mechanical model of the scattering process is also developed and shown to have a good qualitative correspondence to the molecular dynamics results. This model enables us to calculate the phonon transmission at a boundary based on the buckling height and width.

The paper is organized as follows: Section II describes the simulation method and the construction of the phonon wave packets. Section III presents the results of the molecular dynamics simulations, while section IV describes the continuum mechanical treatment of the problem. Finally, section V contains discussion and conclusions. 

\section{Method}
Molecular dynamics (MD) simulations have been performed using the program package LAMMPS (Large-scale Atomic/Molecular Massively Parallel Simulator) \cite{Plimpton1995LAMMPS}, and the interaction between carbon atoms has been modelled using the Tersoff bond-order potential \cite{Tersoff1988Si}. The Tersoff potential is known to reproduce lattice constants and elastic properties of several carbon allotropes  with reasonable accuracy \cite{Tersoff1988C}. In the present study the potential parameters obtained by Lindsay and Broido \cite{Lindsay2010potential} were used, as this parametrization gives an improved description of the phonon dispersion in graphene. The potential gives a lattice parameter of \unit[0.249]{nm}.

\subsection{Grain boundaries}
\begin{figure}
  \begin{center}
    \includegraphics[scale=1]{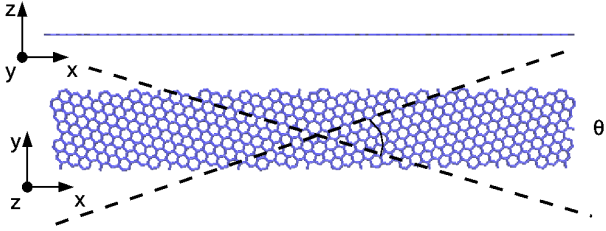}
    \caption{\label{fig:gb32_2} (Color online) Grain boundary with misorientation angle $32.2^\circ$, seen from the $y$ direction (top) and from the $z$ direction (bottom). Figure created using VMD\cite{HUMP96}.}
    \end{center}
\end{figure}

\begin{figure*}
  \begin{center}
    \includegraphics[scale=1]{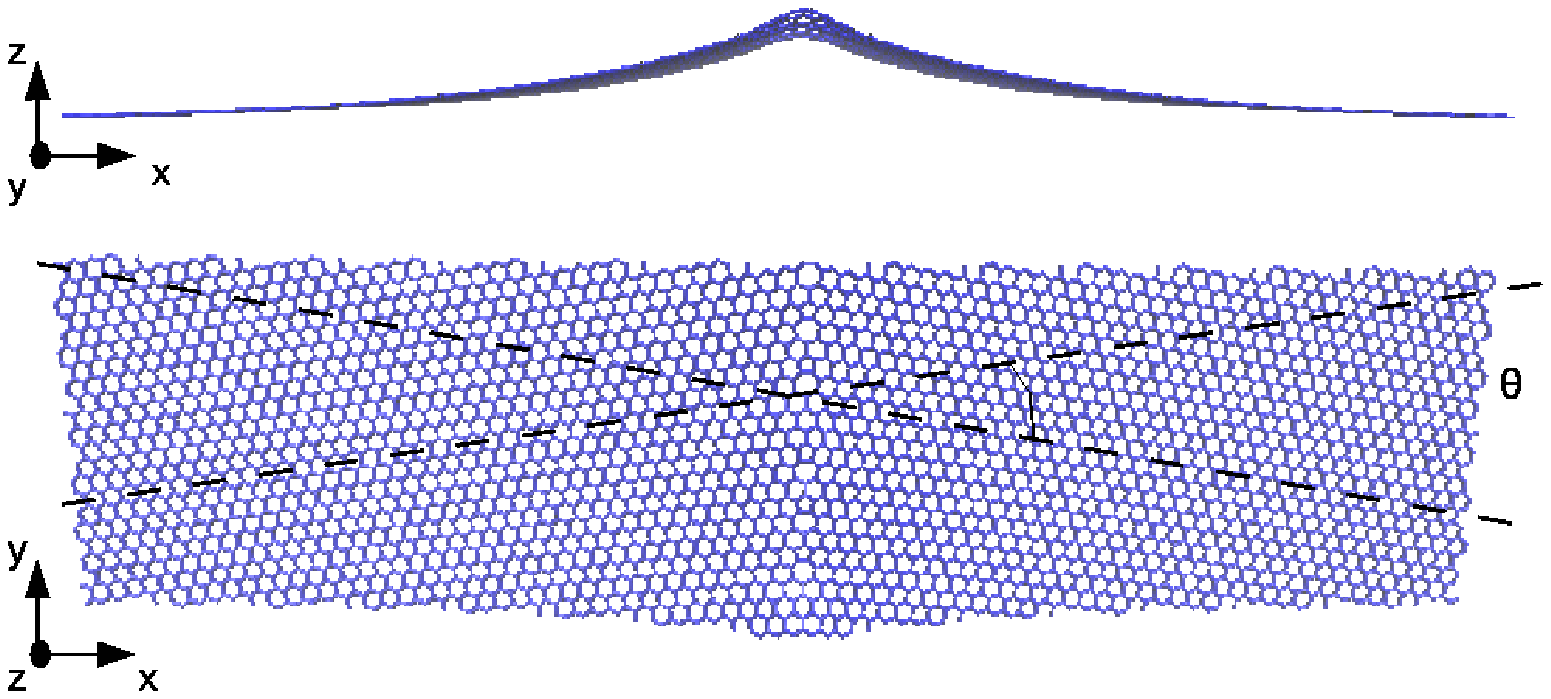}
    \caption{\label{fig:gb17_9}(Color online) Grain boundary with misorientation angle $17.9^\circ$, seen from the $y$ direction (top) and from the $z$ direction (bottom). Figure created using VMD\cite{HUMP96}.}
    \end{center}
\end{figure*}

\begin{figure}
  \begin{center}
    \includegraphics[scale=1]{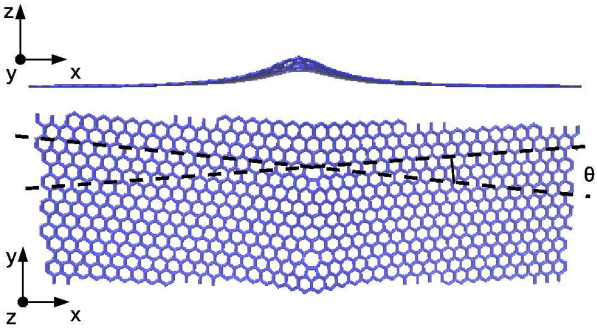}
    \caption{\label{fig:gb9_4}(Color online) Grain boundary with misorientation angle $9.4^\circ$, seen from the $y$ direction (top) and from the $z$ direction (bottom). Figure created using VMD\cite{HUMP96}.}
    \end{center}
\end{figure}

The structure and energetics of graphene grain boundaries have previously been extensively studied using atomistic simulation methods \cite{Carlsson2011gbstruct,Liu2011gbstruct,Yazyev2012energy,Yazyev2010defects,Zhang2013nonsym}. Tilt grain boundaries in graphene have been shown to consist of dislocations in the form of pentagon-heptagon defect pairs, with the defect density determined by the misorientation angle. In particular, Carlsson et al. have shown that it is possible to predict the structure of tilt grain boundaries in graphene using coincidence site lattice theory \cite{Carlsson2011gbstruct}. In the present study, three of the symmetric tilt grain boundaries that were studied by Carlsson et al. are considered. The boundaries are characterized by the misorientation angles $32.2^\circ$, $17.9^\circ$ and $9.4^\circ$ and can be seen in Figures \ref{fig:gb32_2}, \ref{fig:gb17_9} and \ref{fig:gb9_4}, respectively.

 Grain boundaries have been constructed using the Atomistic Simulation Environment\cite{ASE}. To construct a grain boundary, two sheets of graphene are rotated with respect to one another by the desired misorientation angle $\theta$. The grains are then joined and any atoms overlapping with each other are removed. In theory, no extra atoms should have to be added to the structure. In practice, however, the short range of the interatomic potential makes it necessary to add atoms in positions where the nearest neighbors are more than twice the bulk interatomic distance apart. If no atoms are added, the resulting structure will contain under-coordinated atoms. After addition of necessary atoms, the structure is optimized using a conjugate gradients energy minimization method and the resulting structure is checked for under-coordinated atoms. If any under-coordinated atoms are found, more atoms are added and the minimization process repeated. To enable full relaxation, the size of the simulated system in the directions parallel and perpendicular to the grain boundary are allowed to vary independently of each other. Periodic boundary conditions are applied and each simulated system therefore contains two grain boundaries. 

Although several studies find that most graphene grain boundaries are buckled \cite{Carlsson2011gbstruct,Liu2011gbstruct,Yazyev2012energy,Yazyev2010defects,Zhang2013nonsym}, the energy minimization algorithms in LAMMPS do not succeed in producing buckled grain boundaries. Therefore, the optimized grain boundary structure is equilibrated in a NPT simulation at $T= 50$ K and $P=0$ bar for \unit[500]{ps} with a timestep of \unit[1]{fs}, and thereupon cooled to \unit[0.01]{K} at a rate of \unit[6]{K\,ns$^{-1}$}. After this a final energy minimization, in which the shape of the system is allowed to change, is performed. This produces systems with little residual thermal energy and in most cases a substantial buckling. Equilibrating at a higher temperature does not change the grain boundary structure or buckling characteristics.

\subsection{Wave packet method}
In order to obtain detailed information about the phonon scattering it is necessary to introduce phonons with a well-defined polarization and frequency into the system. However, the limitations regarding system size preclude the use of plane waves. We therefore adopt the wave-packet method of Schelling et al. \cite{Schelling2002WP1, Schelling2004WP2, Kimmer2007WP3}, who have used it to study thermal conductivity across grain boundaries in semiconductors. In this approach, phonon wave-packets are constructed from the vibrational eigenmodes of the lattice according to 
\begin{equation}\label{eq:wavepacket}
\mathbf{u}_j=\sum_{\mathbf{k}}a_{\mathbf{k}}\mathbf{\epsilon}_{j\mathbf{k}}e^{i(\mathbf{k}\cdot\mathbf{r}_j-\omega(\mathbf{k})t)},
\end{equation}
where $\mathbf{u}_j$ is a vector describing the displacement of atom $j$, $\mathbf{k}$ is a wavevector, $\mathbf{\epsilon}_{j\mathbf{k}}$ is the polarization vector for the appropriate phonon branch, $\mathbf{r}_j$ is the position of atom $j$ and $\omega$ is the phonon frequency. The amplitudes $a_{\mathbf{k}}$ are given by
\begin{equation}\label{eq:pulseamp}
a_{\mathbf{k}}=A\exp(-\eta^2\vert \mathbf{k}-\mathbf{k}_0\vert^2)\exp(-i\mathbf{k}\cdot\mathbf{R}_0)
\end{equation}
where $A$ is an amplitude and $\eta$ is the width of the wavepacket in real space. The wavepacket is centered around $\mathbf{k}_0$ in reciprocal space and around $\mathbf{R}_0$ in real space. Initial velocities can be obtained by differentiating Eq. (\ref{eq:wavepacket}) with respect to time. 

The polarization vectors $\mathbf{\epsilon}_{j\mathbf{k}}$ and dispersion relation $\omega(\mathbf{k})$ have been obtained by diagonalizing the dynamical matrix of the perfect lattice using the General Utility Lattice Program (GULP) \cite{Gale1997GULP1, Gale2003GULP2}. For the wavepacket width $\eta$ a value of \unit[5]{nm} was chosen, in order to make the wavepackets narrow in reciprocal space and thereby reduce the distortion of the pulse with time. This is important due to the quadratic dispersion of the out-of-plane phonon mode in graphene. However, as this makes the wavepackets quite wide in real space the simulations require systems that are at least a few hundred nanometers long in the direction of propagation (perpendicular to the grain boundary). The amplitude was chosen to be small, \unit[0.013]{nm}, in order to avoid nonlinear effects\cite{Atalaya2008nonlin,Eriksson2013circular}. All wave-packet simulations are conducted with periodic boundary conditions in the direction parallel to the grain boundary and fixed boundary conditions in the other directions.

\section{Results}
\subsection{Grain boundaries}
The grain boundary energy $\gamma$ of each boundary has been calculated according to
\begin{equation}\label{eq:gamma}
\gamma=\frac{E_{\mathrm{GB}}-NE_{\mathrm{0}}}{2L},
\end{equation}
where $E_{\mathrm{GB}}$ is the energy of a simulated grain boundary system, $N$ is the number of atoms in the system, $E_0$ is the energy per atom of the perfect lattice and $L$ is the length of the simulated system in the direction parallel to the boundary. The factor of $1/2$ appears because there are two grain boundaries in each system due to periodic boundary conditions. 

Both the $32.2^\circ$ and the $9.4^\circ$ grain boundary have grain boundary energies of \unit[3.7]{eV\,nm$^{-1}$}, while the $17.9^\circ$ grain boundary has an energy of \unit[4.6]{eV\,nm$^{-1}$}. These values are similar to those obtained by Liu et al. \cite{Liu2011gbstruct} using the AIREBO potential, which is closely related to the Tersoff potential. Both potentials overestimate the grain boundary energies slightly compared to calculations using density functional theory or more long-ranged potentials \cite{Carlsson2011gbstruct,Zhang2013nonsym,Yazyev2012energy}. For example, Carlsson et al. report grain boundary energies about $1$ to \unit[1.5]{eV\,nm$^{-1}$} lower \cite{Carlsson2011gbstruct}. The grain boundary energies found in the present study can be seen in Table \ref{tab:gbenergies}.

As mentioned in the Methods section, out-of-plane buckling is an important characteristic of graphene grain boundaries. One of the boundaries considered here, the $32.2^\circ$ grain boundary, is found to be flat in agreement with previous studies \cite{Carlsson2011gbstruct,Liu2011gbstruct}. The other two, the $9.4^\circ$ and $17.9^\circ$ boundaries, display grain boundary buckling in the shape of a ridge running along the boundary. Due to the distribution of defects in the boundary region the height of the ridge varies along the grain boundary. The buckling direction (upwards or downwards from the graphene sheet) appears to be random, and the two grain boundaries present in the same simulated system are not necessarily buckled in the same direction. 

To characterize the buckling we have estimated a buckling height and width. The buckling height $H$ should be measured relative to the unperturbed graphene sheet far from the boundary. However, in systems that have been subjected to nonzero temperatures the graphene sheet does not become perfectly flat even far from the boundaries. Therefore, a mean value of the position in the out-of-plane ($z$) direction is calculated for a region of the sheet located as far away from both grain boundaries as possible and about \unit[10]{nm} wide in the direction perpendicular to the boundary ($x$). The buckling height is taken to be the difference between this reference value and the peak (upwards or downwards) of the boundary buckle. The buckling width $W$ is estimated as the width of the buckle halfway between the peak and the reference value. 

\begin{table}
\centering
\caption{\label{tab:gbenergies} Grain boundary energies. The misorientation angle is denoted by $\theta$ and the boundary energy by $\gamma$. Grain boundary energies obtained by Liu et al. (Figure 4 in Ref. \onlinecite{Liu2011gbstruct}) are included for comparison.}
\begin{ruledtabular}
\begin{tabular}{r c c }
                 $\theta$ ($^\circ$) & $\gamma$ (eV\,nm$^{-1}$) & $\gamma$ from Ref. \onlinecite{Liu2011gbstruct} (eV\,nm$^{-1}$)  
                 \\ \hline
                 32.2 & 3.7 & 3.9
 \\ 
                 17.9 & 4.6 & 4.6 
 \\ 
                 9.4  & 3.7 & 3.5 
\\
\end{tabular}
\end{ruledtabular}
\end{table}

\begin{figure}
  \begin{center}
    \includegraphics[scale=1]{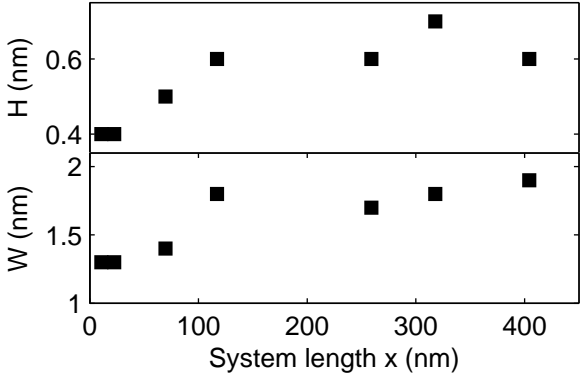}
    \caption{\label{fig:buckle_9_4} Buckling height (top) and width (bottom) for the $9.4^\circ$ boundary plotted against the length of the simulated system in the $x$ direction.}
    \end{center}
\end{figure}

\begin{figure}
  \begin{center}
    \includegraphics[scale=1]{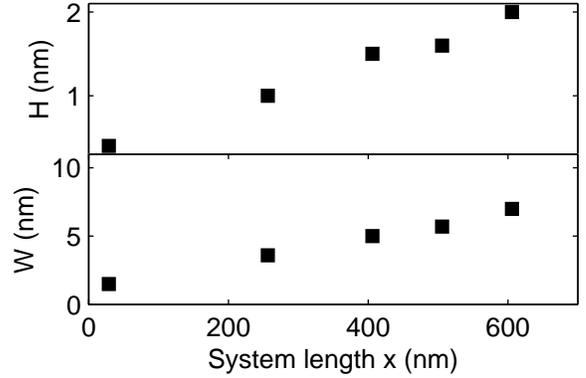}
    \caption{\label{fig:buckle_17_9} Buckling height (top) and width (bottom) for the $17.9^\circ$ boundary plotted against the length of the simulated system in the $x$ direction.}
    \end{center}
\end{figure}

In order to find the buckling characteristics of the $9.4^\circ$ and $17.9^\circ$ grain boundaries in the limit of large grains, systems of different sizes have been investigated. It is found that increasing the system size in the $y$ direction to more than one grain boundary period does not affect the buckling characteristics. The system length in the $x$ direction, however, has a considerable effect on the buckling as can be seen in Figures \ref{fig:buckle_9_4} and \ref{fig:buckle_17_9}. The buckling height of the $9.4^\circ$ grain boundary increases with increasing length for systems smaller than \unit[100]{nm} and then appears to fluctuate around \unit[0.65]{nm} for systems between $100$ and \unit[400]{nm} in length. Similarly, the buckling width fluctuates around \unit[1.8]{nm} for systems longer than \unit[100]{nm}. For the $17.9^\circ$ boundary, the buckling height and width are seen to grow for system sizes up to \unit[600]{nm}. The rise in computational cost with increasing system size regrettably prevents us from extending the study to even larger systems. Still, it is evident from these results that the buckling characteristics converge very slowly with respect to system length in the $x$ direction.

 It should be noted that the increase in buckling height and width with increasing system size is accompanied by surprisingly small changes in the grain boundary energy. For example, the grain boundary energy of the $9.4^\circ$ boundary obtained with a cell \unit[22.5]{nm} long is \unit[3.8]{eV\,nm$^{-1}$}, only \unit[0.1]{eV\,nm$^{-1}$} higher than that obtained with a cell \unit[259.1]{nm} long.

Even without complete convergence with respect to system size it is clear that the present study finds significantly larger buckling heights compared to previous work. For example, although Carlsson et al.\cite{Carlsson2011gbstruct} find a height close to \unit[0.5]{nm} for the $9.4^\circ$ boundary using first-principles methods, they obtain a lower buckling height for the $17.9^\circ$ boundary (about \unit[0.3]{nm}). Similarly, Liu et al.\cite{Liu2011gbstruct} find the height of the $9.4^\circ$ grain boundary to be about \unit[0.18]{nm} and the height of the $17.9^\circ$ boundary to be slightly smaller, about \unit[0.16]{nm}. The most probable cause of this discrepancy is that the previous studies consider fairly small systems. Carlsson et al.\cite{Carlsson2011gbstruct} state that the grain boundaries in their periodic simulation cells are separated by twice the grain boundary period, which in the case of the $9.4^\circ$ boundary implies a distance of \unit[3.03]{nm} and for the $17.9^\circ$ boundary a distance of \unit[4.808]{nm}. Liu et al.\cite{Liu2011gbstruct} are regrettably vague about the separation between grain boundaries but mention minimum grain sizes of \unit[3.6]{nm}.

The grain boundary buckling height is also influenced by the boundary conditions, as has been seen for crystalline membranes\cite{Carraro1993membrane}, and by whether or not the system size is allowed to change. In the present study we use periodic boundary conditions during grain boundary fabrication, and allow the size of the system to change isotropically during the heating and cooling process. A fixed size would have prevented the system from contracting in the direction perpendicular to the boundary, producing a lower buckling. Similarly, an even larger buckling would have been obtained by allowing the system to change size anisotropically during the NPT simulation, enabling it to contract in the direction perpendicular to the grain boundary while keeping the size in the parallel direction constant. Which of these approaches would be optimal for reproducing the experimental behavior of suspended polycrystalline graphene depends on the fabrication process. However, the approach used here produces grain boundary systems adequate for the phonon scattering simulations that are the primary focus of this study.

\subsection{Phonon scattering}
\begin{table}
\centering
\caption{\label{tab:boundaries} Grain boundary buckling and simulation cell dimensions. The misorientation angle is denoted by $\theta$, the buckling height by $H$ and the buckling width by $W$. The buckling height obtained by Liu et al. (Figure 5 in Ref. \onlinecite{Liu2011gbstruct}) is included for comparison.}
\begin{ruledtabular}
\begin{tabular}{r c c c c }
                 $\theta$ ($^\circ$) & Dimensions (nm$^3$) & $H$ (nm) & $H$, Ref. \onlinecite{Liu2011gbstruct} (nm) & $W$ (nm)  
                 \\ \hline
                 32.2 & $264\times0.98\times10$  & 0   & 0 & 0
\\ 
                 17.9 & $406\times2.404\times10$ & 1.5 & 0.16 & 5
\\ 
                 9.4  & $259\times1.516\times10$ & 0.6 & 0.18 & 1.7
\\
\end{tabular}
\end{ruledtabular}
\end{table}

For the phonon scattering simulations we have chosen systems with a size of one grain boundary period in the $y$ direction. The systems containing the $32.2^\circ$ and $9.4^\circ$ boundaries are approximately \unit[250]{nm} long in the $x$ direction, while the system containing the $17.9^\circ$ boundary is about \unit[400]{nm} long. This system length has been chosen so as to give the 17.9$^\circ$ boundary a substantially higher buckling than the $9.4^\circ$ boundary, since the buckling characteristics of the $17.9^\circ$ boundary show no signs of convergence with respect to system size. System sizes and buckling characteristics for each boundary can be seen in Table \ref{tab:boundaries}. 

\begin{figure}
  \begin{center}
    \includegraphics[scale=1]{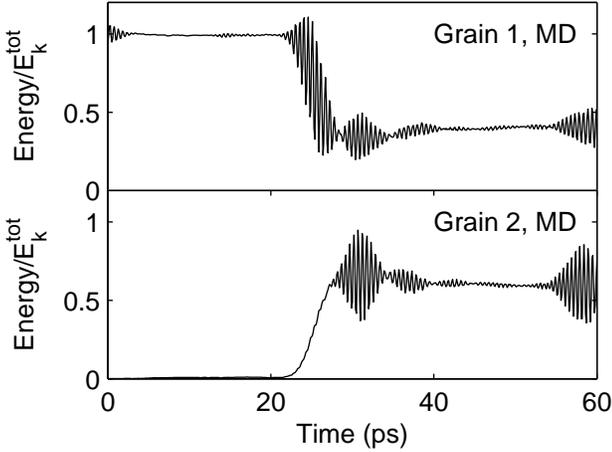}
    \caption{\label{fig:Ekin_regions}Kinetic energy in grain 1 (top) and grain 2 (bottom) as a function of time, for the $9.4^\circ$ boundary with $k_0=3~\mathrm{nm^{-1}}$.}
    \end{center}
\end{figure}

Figure \ref{fig:Ekin_regions} shows the fraction of the average total kinetic energy on either side of the grain boundary as a function of time for a wavepacket with $k_0=3$ nm$^{-1}$ incident on the $9.4^\circ$ boundary. Initially, all kinetic energy is located in grain 1. After about \unit[20]{ps} the kinetic energy in grain 1 starts to decrease and the kinetic energy in grain 2 starts to increase, indicating that the wavepacket is scattered against the boundary. At \unit[35]{ps} the energy in grain 2 decreases by about \unit[9]{\%} and the energy in grain 1 increases by \unit[17]{\%}, which is rather unexpected as the phonon pulse should have passed the boundary completely at that time. Between $40$ and \unit[60]{ps} the energy in each grain fluctuates around a constant value. The large fluctuations at \unit[60]{ps} indicate interaction with the fixed boundary conditions.

\begin{figure}
  \begin{center}
    \includegraphics[scale=1]{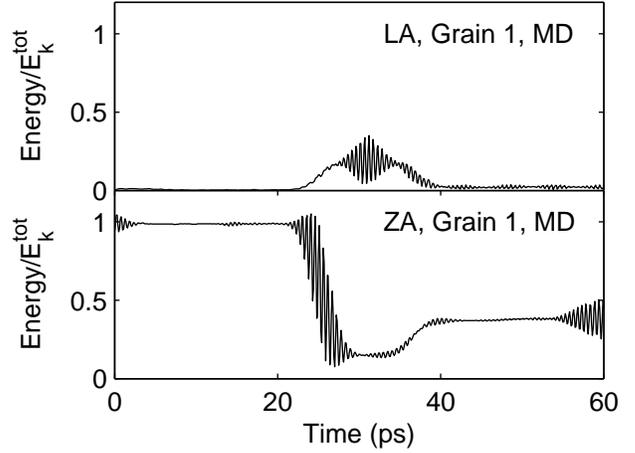}
    \caption{\label{fig:Ekin_d1} Fraction of the total kinetic energy in the longitudinal mode (top) and the out-of-plane mode (bottom) in grain 1, for the $9.4^\circ$ boundary with $k_0=3~\mathrm{nm}^{-1}$.}
    \end{center}
\end{figure}

\begin{figure}
  \begin{center}
    \includegraphics[scale=1]{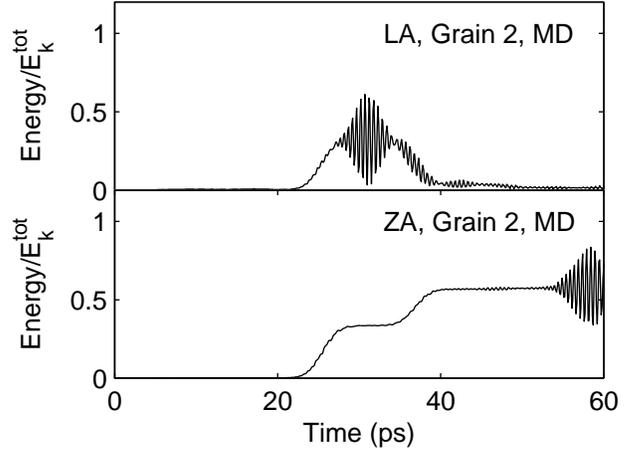}
    \caption{\label{fig:Ekin_d2}Fraction of the total kinetic energy in the longitudinal mode (top) and the out-of-plane mode (bottom) in grain 2, for the $9.4^\circ$ boundary with $k_0=3~\mathrm{nm^{-1}}$.}
    \end{center}
\end{figure}

To investigate the unexpected behaviour at \unit[35]{ps} the contributions to the kinetic energy from movement in different directions is considered, as can be seen in Figures \ref{fig:Ekin_d1} and \ref{fig:Ekin_d2}. As expected, all kinetic energy is initially due to movement in the $z$ direction (the out-of-plane mode, or ZA mode). As the pulse hits the boundary at \unit[20]{ps}, a contribution to kinetic energy due to movement in the $x$ direction (longitudinal mode, or LA mode) appears on both sides of the boundary. After \unit[35]{ps} the movement in the $x$ direction starts to decrease, and at \unit[40]{ps} almost all the kinetic energy is again in the out-of-plane mode. No movement in the $y$ direction (in-plane and parallel to the grain boundary) is seen during the simulation.

The behavior of the longitudinal vibrations can be understood by considering the propagation velocity of the vibrations. According to the dispersion relation of longitudinal acoustic phonons in graphene, the propagation velocity should approach \unit[22]{nm\,ps$^{-1}$} at long wavelenghs \cite{Lindsay2010potential}. Long-wavelength flexural vibrations are much slower, with a propagation velocity of about \unit[3.7]{nm\,ps$^{-1}$} at $k_0=3$~nm$^{-1}$. Since the simulated system is \unit[259]{nm} long in the $x$ direction and the grain boundary is located in the middle, it takes about \unit[6]{ps} for longitudinal vibrations arising at the grain boundary to reach the end of the system. The vibrations will be reflected against the fixed boundary conditions and return to the grain boundary after an additional \unit[6]{ps}. This corresponds well to the time during which the longitudinal vibrations are seen in the simulation, especially if it is taken into account that the pulse of longitudinal vibrations will have a finite width in space.

 What is seen in Figures \ref{fig:Ekin_d1} and \ref{fig:Ekin_d2} is thus that the scattering of out-of-plane vibrations against the grain boundary produces longitudinal vibrations, which propagate to the edge of the cell where they are reflected. When the reflected longitudinal vibrations again reach the grain boundary they are scattered back into the out-of-plane mode, explaining the decrease in longitudinal vibrations and increase in out-of-plane vibrations between $35$ and \unit[40]{ps}. 

It follows from the above discussion that the changes occurring in the kinetic energy of the two grains after scattering of the longitudinal vibrations result from interactions with the boundary conditions. Transmission and reflection coefficients $T$ and $R$, which are defined as
\begin{align}\label{eq:TR}
\nonumber T=\frac{\langle E_{\mathrm{k}}^{\mathrm{Grain 2}}\rangle}{E_{\mathrm{k}}^{\mathrm{tot}}},\\
\\
\nonumber R=\frac{\langle E_{\mathrm{k}}^{\mathrm{Grain 1}}\rangle}{E_{\mathrm{k}}^{\mathrm{tot}}},
\end{align}
must therefore be evaluated before this point. Here, $E_{\mathrm{k}}^{\mathrm{Grain 1}}$ and $E_{\mathrm{k}}^{\mathrm{Grain 2}}$ are the time-dependent kinetic energies in grain 1 and grain 2, $E_{\mathrm{k}}^{\mathrm{tot}}$ is the average kinetic energy of the entire system and the brackets represent a time average over times between the scattering of the incident pulse at the grain boundary and the scattering of the reflected longitudinal vibrations.

\begin{figure}
  \begin{center}
    \includegraphics[scale=1]{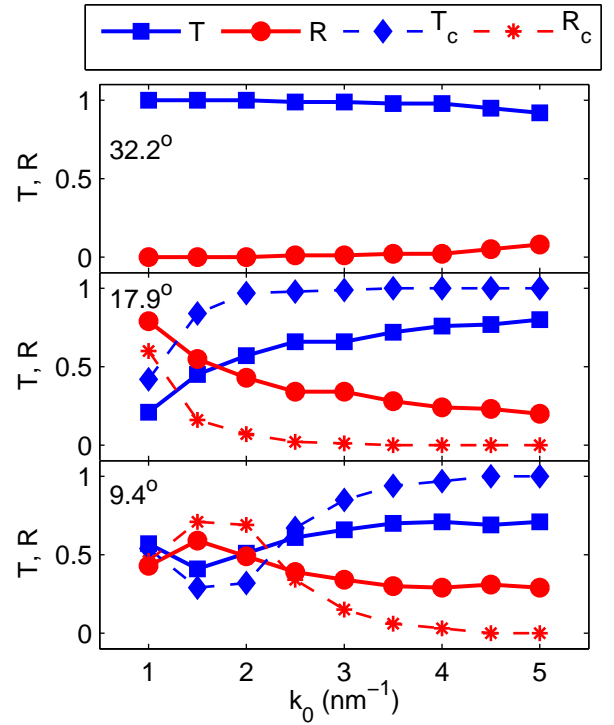}
    \caption{\label{fig:TRall}(Color online) Energy transmission coefficient $T$ and reflection coefficient $R$ as a function of $k_0$ for the $32.2^\circ$ grain boundary (top), the $17.9^\circ$ boundary (middle) and the $9.4^\circ$ boundary (bottom). $T_{\mathrm{c}}$ and $R_{\mathrm{c}}$ are the corresponding results from the continuum mechanical model described in Section \ref{sec:continuum}. The continuum model parameters are $\chi=$ \unit[1.5]{nm}, FWHM=\unit[5]{nm} for the $17.9^\circ$ boundary and $\chi=$ \unit[0.6]{nm}, FWHM=\unit[1.7]{nm} for the $9.4^\circ$ boundary.}
    \end{center}
\end{figure}

Figure \ref{fig:TRall} shows $T$ and $R$ as functions of $k_0$ for all three grain boundaries. It is clear that the $32.2^\circ$ boundary, which is flat, has $T\approx1$ and $R\approx0$ for the entire range of wavenumbers. Thus, this boundary appears not to scatter long-wavelength out-of-plane vibrations. In contrast, the transmission at the buckled $17.9^\circ$ boundary is as low as $T=0.2$ for $k_0=1$ nm$^{-1}$ and has a maximum value of $T=0.8$ at $k_0=5$ nm$^{-1}$, showing that it scatters the incoming pulse. The $9.4^\circ$ boundary also causes significant scattering as the transmission never exceeds $T=0.7$ and reaches a minimum value of $T=0.4$ at $k_0=1.5$ nm$^{-1}$. 

The differences in transmission at the flat and buckled boundaries suggest that it is the buckling of the grain boundary that scatters long-wavelength vibrations. This is reasonable considering that the lattice defects present in all three grain boundaries are only between $0.2$ and \unit[0.3]{nm} wide, while the buckling of the $17.9^\circ$ boundary is \unit[5]{nm} wide and that of the $9.4^\circ$ \unit[1.7]{nm} wide. Vibrations with wavelengths above \unit[1]{nm}, such as those studied here, are unlikely to be strongly affected by the small lattice defects but will interact with the wider buckling. 

To provide a measure of how much of the energy is scattered into the longitudinal mode, the longitudinal contributions to the transmission ($T_{\mathrm{L}}$) and reflection ($R_{\mathrm{L}}$) are calculated as
\begin{align}
\nonumber T_{\mathrm{L}}=\frac{\left\langle\sum_{\mathrm{Grain 2}} m_{\mathrm{C}}v_{i\mathrm{x}}^2/2\right\rangle}{E_{\mathrm{k}}^{\mathrm{tot}}}\\
\\
\nonumber R_{\mathrm{L}}=\frac{\left\langle\sum_{\mathrm{Grain 1}} m_{\mathrm{C}}v_{i\mathrm{x}}^2/2\right\rangle}{E_{\mathrm{k}}^{\mathrm{tot}}}
\end{align}
where $v_{i,\mathrm{x}}$ is the velocity of atom $i$ in the $x$ direction, $m_{\mathrm{C}}$ is the mass of a carbon atom and the summations run over all atoms in one grain. As in Equation \ref{eq:TR}, the brackets represent a time average over times between the scattering of the incident pulse at the grain boundary and the scattering of the reflected longitudinal vibrations.

\begin{figure}
  \begin{center}
    \includegraphics[scale=1]{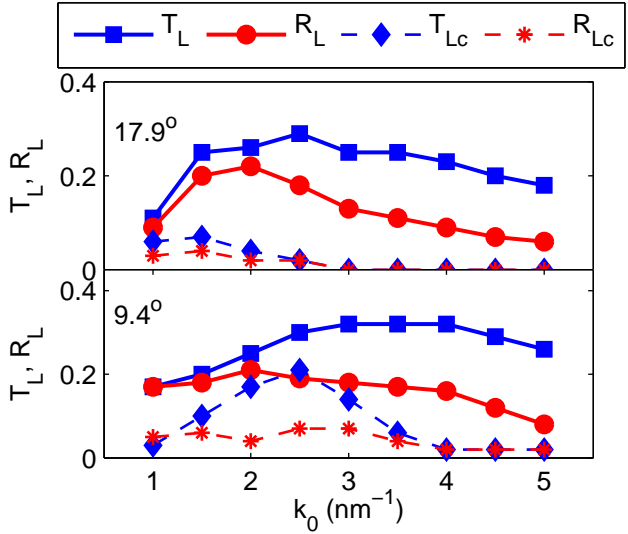}
    \caption{\label{fig:TRLall}(Color online) Energy transmission $T_{\mathrm{L}}$ and reflection $R_{\mathrm{L}}$ to the longitudinal mode as a function of $k_0$ for the $17.9^\circ$ grain boundary (top) and the $9.4^\circ$ grain boundary (bottom). $T_{\mathrm{Lc}}$ and $R_{\mathrm{Lc}}$ are corresponding results from the continuum mechanical model described in Section \ref{sec:continuum}. The continuum model parameters are $\chi=$ \unit[1.5]{nm}, FWHM=\unit[5]{nm} for the $17.9^\circ$ boundary and $\chi=$ \unit[0.6]{nm}, FWHM=\unit[1.7]{nm} for the $9.4^\circ$ boundary.} 
    \end{center}
\end{figure}

Figure \ref{fig:TRLall} shows $T_{\mathrm{L}}$ and $R_{\mathrm{L}}$ as a function of $k_0$ for the $17.9^\circ$ and $9.4^\circ$ grain boundaries. It appears that $T_{\mathrm{L}}>R_{\mathrm{L}}$ for all $k_0$ at both grain boundaries. Also, the maximum value of $T_{\mathrm{L}}$ occurs at a higher wavenumber, i.e. a shorter wavelength, for the $9.4^\circ$ boundary than for the $17.9^\circ$ boundary.

\section{Continuum mechanical model}
\label{sec:continuum}
\subsection{Theory}
To analyze the scattering process a continuum mechanical model of the system has been constructed. Continuum mechanics is valid as long as variations in the deformation of the graphene sheet occur on a length scale larger than the lattice parameter. This is mostly true in the present case since the lattice parameter of graphene is \unit[0.24]{nm} and the wavelengths considered are above \unit[1]{nm}. Since no movement in the $y$ direction was seen in the MD simulations, the system is modeled as a string restricted to vibrating in one transversal mode and the longitudinal mode. The equation governing the propagation of a transversal displacement $w(x,t)$ is
\begin{equation}\label{eq:wavew}
\rho\ddot{w}+\kappa\partial_x^4w-\partial_x\sigma_{xx}(\partial_xw)=0
\end{equation}
where $\rho$ is the density and $\kappa$ the bending rigidity, $\partial_x$ denotes derivation with respect to $x$ and the stress is given by
\begin{equation}\label{eq:sigmaxx}
\sigma_{xx}=\sigma_0+(\lambda+2\mu)\left(\partial_x u+\frac{1}{2}(\partial_xw)^2\right).
\end{equation}
In this expression $\sigma_0$ is a preexisting stress in the string, $\lambda$ and $\mu$ are Lam\'e parameters and $u$ is the longitudinal displacement which obeys the equation
\begin{equation}\label{eq:waveu}
\rho\ddot{u}-\partial_x\sigma_{xx}=0.
\end{equation}

The buckling can be introduced as time-independent terms in $w(x,t)$ and $u(x,t)$:
\begin{align}
\nonumber w(x,t)&=w_0(x)+w_1(x,t),\\
&\\
\nonumber u(x,t)&=u_0(x)+u_1(x,t).
\end{align}
Since the amplitudes of the time-dependent vibrations are small, we can omit terms that are nonlinear in the derivatives of $w_1(x,t)$ and $u_1(x,t)$. Inserting the expression for the stress into Equation (\ref{eq:wavew}) then gives
\begin{align}\label{eq:wavew_leading}
\nonumber&\rho\ddot{w}_1+\kappa\partial_x^4w_1-\sigma_0\partial_x^2w_1-\\
\nonumber&(\lambda+2\mu)\partial_x\left(\partial_x u_1+\partial_xw_0(x)\partial_xw_1\right)\partial_xw_0(x)-\\
&(\lambda+2\mu)\partial_x\left(\partial_x u_0(x)+\frac{1}{2}(\partial_xw_0(x))^2\right)\partial_xw_1=0.
\end{align}
Treating Equation (\ref{eq:waveu}) in the same manner we obtain
\begin{equation}\label{eq:waveu_leading}
\rho\ddot{u}_1-(\lambda+2\mu)\partial_x^2u_1-(\lambda+2\mu)\partial_x(\partial_xw_0\partial_xw_1)=0.
\end{equation}

\subsection{Method of solution}
Numerical solutions of Equations \ref{eq:wavew_leading} and \ref{eq:waveu_leading} have been obtained using finite-difference methods. The equations have been discretized using standard discretization schemes \cite{numrec} and the time-dependent displacements $w_1(x,t)$ and $u_1(x,t)$ have been calculated at discrete positions $x_i=i\Delta x$ and times $t_n=n\Delta t$, with step sizes $\Delta x=0.05$ nm and $\Delta t=0.4\sqrt{dx^4/4\kappa}=0.8$ fs. The timestep has been chosen in accordance with the stability criterion for the Euler-Bernoulli equation \cite{Tzes1989EB} and is calculated using the value of the bending rigidity given by the modified Tersoff potential ($\kappa$=\unit[$2.8\times10^{-19}$]{J}). Also for $\rho$, $\lambda$ and $\mu$ the values given by the interatomic potential have been used, i.e. $\rho=7.42\times10^{-7}$ kg\,m$^{-2}$ and $\lambda+2\mu=356$ N\,m$^{-1}$. The preexisting stress $\sigma_0$ has been used as a fitting parameter in order to obtain the same propagation velocities as in the MD simulations, giving it a value of $5\times10^{-3}\times(\lambda+2\mu)=1.78$ N\,m$^{-1}$. 

Fixed boundary conditions are applied in the $x$ direction. A pulse of transverse vibrations, similar to the wave-packets used in the MD simulations, is introduced through the initial conditions:
\begin{align}\label{eq:wave_continuum}
\nonumber&w_1(x_i,0)=A_ce^{ik_0x_i}e^{-(x_i-3L/8)^2/\sigma^2} \\
\\
\nonumber&\partial_tw_1(x_i,0)=\partial_t\left(A_ce^{ik_0x_i-i\omega(k_0)t_n}e^{-(x_i-3L/8)^2/\sigma^2}\right),
\end{align}
where $A_c=0.01$ nm is the vibration amplitude, $k_0$ is a central wavevector, $\omega(k_0)$ is the vibration frequency at wavevector $k_0$ for the case of zero stress, $\sigma=9$ nm determines the width of the wavepacket and $L=400$ nm is the system length. Note that the simulated string extends from $-L/2$ to $L/2$, with $x=0$ in the middle of the cell. For simplicity the buckling is approximated as a Gaussian function:
\begin{align}\label{eq:buckle_continuum}
\nonumber&w_0(x_i)=\chi e^{-x_i^2/2\xi^2}, \\
&\\
\nonumber&u_0(x_i)=\int dx~\left(-\frac{1}{2}(\partial_xw_0(x_i))^2\right).
\end{align}
The parameters $\chi$ and $\xi$ can be adjusted to mimic boundaries with different buckling height and width. To obtain the longitudinal displacement $u_0(x)$ Equation (\ref{eq:waveu}) is used, with $\sigma_{xx}$ given by Equation (\ref{eq:sigmaxx}) and $w_1(x,t)=0$. This gives the static longitudinal displacement resulting from the static transversal displacement $w_0(x)$.

\subsection{Results}
To compare the results of the continuum mechanical model to those of the MD simulations it is necessary to find the fraction of the total energy in each vibrational mode at either side of the scattering center as a function of time. Assuming that the vibrations are harmonic waves the total kinetic energy is given by
\begin{equation}
E^{\mathrm{tot}}=\frac{\Delta x}{2}\rho\sum_{x_i=-L/2}^{x_i=L/2}\omega_{\mathrm{T}}^2 w_1^2(x_i,t_n)+\omega_{\mathrm{L}}^2u_1^2(x_i,t_n),
\end{equation}
 where $\omega_T$ and $\omega_L$ are the frequencies of the transverse and longitudinal vibrations, respectively. The fraction of the total kinetic energy in each mode is then
\begin{align}
\nonumber \alpha_{\mathrm{T}}(t_n)=\frac{\Delta x\rho\omega^2_{\mathrm{T}}\sum_{x_i>0}w_1^2(x_i,t_n)}{2E^{\mathrm{tot}}}\\
\nonumber\\ 
\nonumber \beta_{\mathrm{T}}(t_n)=\frac{\Delta x\rho\omega_{\mathrm{T}}^2\sum_{x_i<0}w_1^2(x_i,t_n)}{2E^{\mathrm{tot}}} \\
\\ \nonumber
\alpha_{\mathrm{L}}(t_n)=\frac{\Delta x\rho\omega_{\mathrm{L}}^2\sum_{x_i>0}u_1^2(x_i,t_n)}{2E^{\mathrm{tot}}} \\
\nonumber \\ \nonumber
\beta_{\mathrm{L}}(t_n)=\frac{\Delta x\rho\omega_{\mathrm{L}}^2\sum_{x_i<0}u_1^2(x_i,t_n)}{2E^{\mathrm{tot}}},
\end{align}
where $\alpha_{\mathrm{T}}$ is the energy in the transverse mode at $x>0$, $\beta_{\mathrm{T}}$ is the energy in the transverse mode at $x<0$, $\alpha_{\mathrm{L}}$ is the energy in the longitudinal mode at $x>0$ and $\beta_{\mathrm{L}}$ is the energy in the longitudinal mode at $x<0$. Note that due to the narrow range of frequencies in the incident pulse and the neglect of terms nonlinear in $\partial_xw_1(x,t)$, $\omega_{\mathrm{L}}\approx\omega_{\mathrm{T}}$ and the frequencies will cancel in the above expressions.

\begin{figure}
  \begin{center}
    \includegraphics[scale=1]{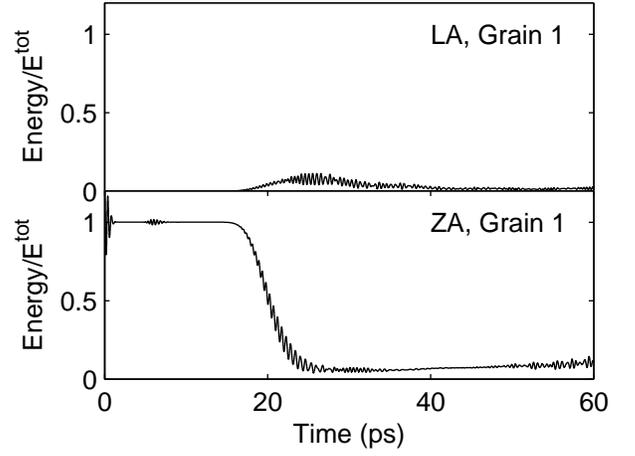}
    \caption{\label{fig:cont_g1} Fraction of the total kinetic energy in the longitudinal mode (top) and the transverse mode (bottom) for $x<0$, obtained from the continuum mechanical model with $\chi=$ \unit[0.6]{nm}, FWHM=\unit[1.7]{nm} and $k_0= $\unit[3]{nm$^{-1}$}.}
    \end{center}
\end{figure}

\begin{figure}
  \begin{center}
    \includegraphics[scale=1]{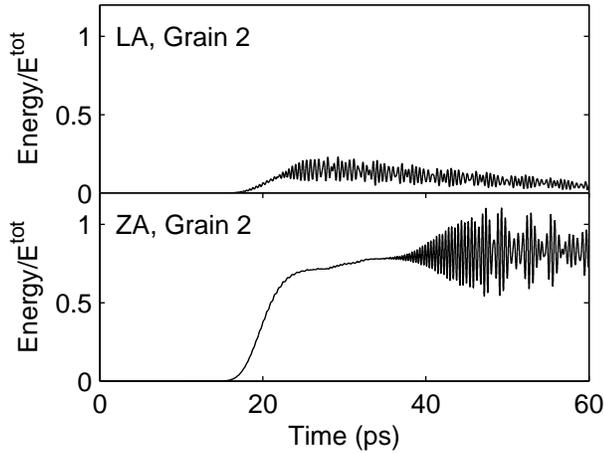}
    \caption{\label{fig:cont_g2} Fraction of the total kinetic energy in the longitudinal mode (top) and the transverse mode (bottom) for $x>0$, obtained from the continuum mechanical model with $\chi=$ \unit[0.6]{nm}, FWHM=\unit[1.7]{nm} and $k_0= $\unit[3]{nm$^{-1}$}.}
    \end{center}
\end{figure}

Figures \ref{fig:cont_g1} and \ref{fig:cont_g2} show $\alpha_{\mathrm{T}}(t_n),~\beta_{\mathrm{T}}(t_n),~\alpha_{\mathrm{L}}(t_n)$ and $\beta_{\mathrm{L}}(t_n)$ for $k_0=3$ nm$^{-1}$. The scattering centre height has been set to $\chi=0.6$ nm and the full width at half maximum (FWHM, $2\xi\sqrt{2\ln(2)}$) to \unit[1.7]{nm} in order to model the $9.4^\circ$ boundary. At the beginning of the simulation all energy is located in the transverse mode at $x<0$. After about \unit[20]{ps} the energy at $x<0$ starts to decrease and the energy at $x>0$ starts to increase. At the same time a longitudinal vibration appears at both $x>0$ and $x<0$. Approximately \unit[25]{ps} into the simulation the amount of energy in the longitudinal mode starts to decrease, while the energy in the transverse mode increases slightly. This behavior is qualitatively similar to the MD results presented in Figures \ref{fig:Ekin_d1} and \ref{fig:Ekin_d2}. The main differences are that the amount of energy transferred to the longitudinal mode is smaller and that it decreases more gradually, which in turn leads to a less abrupt increase of the energy in the transverse mode after scattering of the longitudinal vibration.

As in the MD simulations, transmission and reflection coefficients need to be calculated before any interaction with the boundary conditions. The total transmission and reflection coefficients, $T_{\mathrm{c}}$ and $R_{\mathrm{c}}$, have been calculated according to
\begin{align}
\nonumber T_{\mathrm{c}}=\langle\alpha_{\mathrm{T}}+\alpha_{\mathrm{L}}\rangle \\ 
\\
\nonumber R_{\mathrm{c}}=\langle\beta_{\mathrm{T}}+\beta_{\mathrm{L}}\rangle 
\end{align}
where the brackets denote time averaging over times between the scattering of the incident pulse and the scattering of the reflected longitudinal pulse. 

Figure \ref{fig:TRall} shows $T_{\mathrm{c}}$ and $R_{\mathrm{c}}$ as functions of $k_0$ for values of $\chi$ and FWHM corresponding to the buckling characteristics of the two buckled grain boundaries. Considering first the $9.4^\circ$ boundary, we see that the simple continuum mechanical model with $\chi=0.6$ nm and FWHM = \unit[1.7]{nm} agrees remarkably well with MD results for smaller wavenumbers, even reproducing the minimum in transmission at $k_0=1.5$ nm$^{-1}$. The transmission at $k_0=2$ nm$^{-1}$ is somewhat underestimated, while above $k_0=3$ nm$^{-1}$ it is overestimated compared to MD results. 

\begin{figure}
  \begin{center}
    \includegraphics[scale=1]{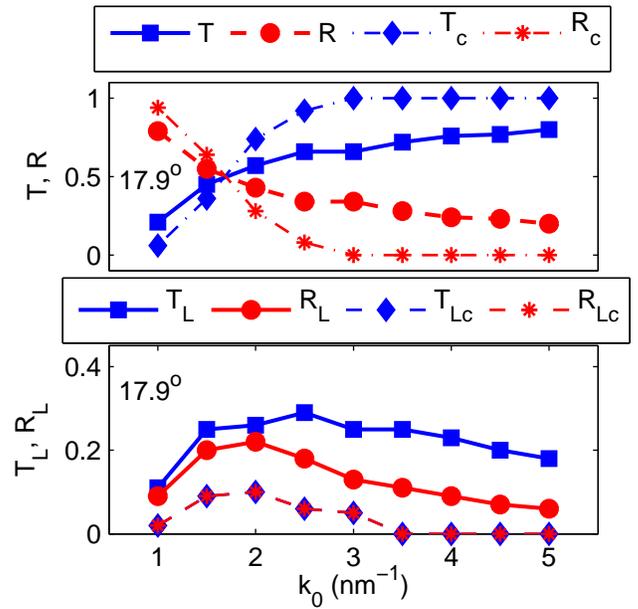}
    \caption{\label{fig:TR17_FWHM20}(Color online) Transmission and reflection coefficients from MD results for the $17.9^\circ$ boundary and corresponding results from the continuum model with $\chi=1.5$ nm and FWHM$=2$ nm. Top: Total energy transmission $T$ and reflection $R$ from MD simulations as a function of $k_0$, with $T_{\mathrm{c}}$ and $R_{\mathrm{c}}$ from the continuum model. Bottom: Energy transmission $T_{\mathrm{L}}$ and reflection $R_{\mathrm{L}}$ to the longitudinal mode from MD simulations as a function of $k_0$, with $T_{\mathrm{Lc}}$ and $R_{\mathrm{Lc}}$ from the continuum model.}
    \end{center}
\end{figure}

For the $17.9^\circ$ boundary there is less agreement between the two models. The continuum mechanical model with $\chi=1.5$ nm and FWHM = \unit[5]{nm} correctly gives $T_{\mathrm{c}}<R_{\mathrm{c}}$ at $k_0=1$ nm$^{-1}$, but overestimates the transmission for all other wavenumbers. To improve the continuum mechanical description different values of FWHM were tested, with FWHM = \unit[2]{nm} producing the best agreement with MD results. As can be seen in Figure \ref{fig:TR17_FWHM20}, a continuum model with FWHM = \unit[2]{nm} reproduces the MD result that $T_{\mathrm{c}}<R_{\mathrm{c}}$ for both $k_0=1$ and $k_0=1.5$ nm$^{-1}$, but still overestimates the transmission at larger $k_0$. 

To further compare the two models, the transmission and reflection into the longitudinal mode were calculated according to
\begin{align}
\nonumber T_{\mathrm{Lc}}=\langle\alpha_{\mathrm{L}}\rangle \\
\\
\nonumber R_{\mathrm{Lc}}=\langle\beta_{\mathrm{L}}\rangle,
\end{align}
where the brackets represent time averaging between the scattering of the incident pulse and scattering of the reflected longitudinal pulse. As can be seen in Figure \ref{fig:TRLall}, the continuum mechanical model agrees with MD simulations in that $T_{\mathrm{Lc}}>R_{\mathrm{Lc}}$ for all wavenumbers and boundaries. However, the continuum mechanical model consistently underestimates the amount of energy in the longitudinal mode compared to MD. For the $9.4^\circ$ boundary continuum mechanical and MD results follow the same general trend up to $k_0=2.5$ nm$^{-1}$, above which $T_{\mathrm{Lc}}$ decreases significantly while the corresponding MD quantity $T_{\mathrm{L}}$ remains nearly constant. For the  $17.9^\circ$ boundary, the continuum model with $\chi=1.5$ and FWHM$=5$ places the maximum value of both $R_{\mathrm{Lc}}$ and $T_{\mathrm{Lc}}$ at $k_0=1$ nm$^{-1}$, while according to the MD simulations $R_{\mathrm{L}}$ and $T_{\mathrm{L}}$ reach their maxima at $k_0=2$ and $k_0=2.5$ nm$^{-1}$ respectively. Once again a much better correspondence is obtained using FWHM = \unit[2]{nm}, as can be seen in Figure \ref{fig:TR17_FWHM20}. While using this width gives $T_{\mathrm{Lc}}=R_{\mathrm{Lc}}$, the maximum value appears at $k_0=2$ nm$^{-1}$ in good agreement with MD results.  

\begin{figure}
  \begin{center}
    \includegraphics[scale=1]{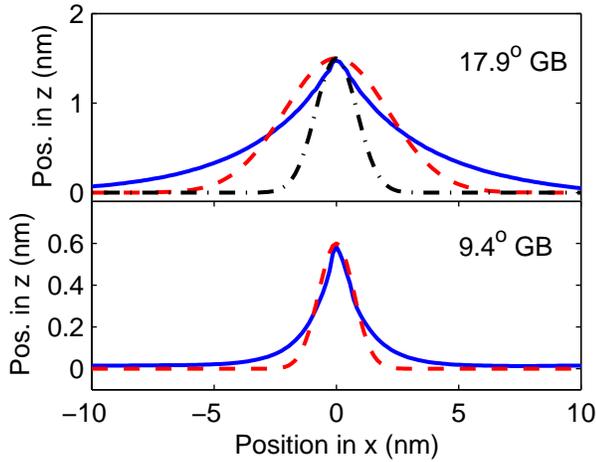}
    \caption{\label{fig:buckling_cf} (Color online) Top: Height profile of the $17.9^\circ$ grain boundary (solid blue line) with $w_0(x)$ for $\chi=1.5$ nm, FWHM=\unit[5]{nm} (dashed red line) and $\chi=1.5$ nm, FWHM=\unit[2]{nm} (dash-dotted black line). Bottom: Height profile of the $9.4^\circ$ grain boundary (solid blue line) with $w_0(x)$ for $\chi=0.6$, FWHM=\unit[1.7]{nm} (dashed red line).}
    \end{center}
\end{figure}

The comparison of the atomistic and continuum mechanical models shows that the scattering at the $9.4^\circ$ boundary can be described by a continuum mechanical model where the Gaussian scattering center has the same height and width as the boundary buckling. In contrast, the scattering at the $17.9^\circ$ boundary is more accurately reproduced by a scattering center narrower than the buckling in width, but with the same height. We attribute this difference to the shape of the boundary buckling. Figure \ref{fig:buckling_cf} shows the height profiles (maximum position in the $z$ direction as a function of position in $x$) for the two boundaries along with the shapes of the scattering centers used in the continuum mechanical model. The shape of the $9.4^\circ$ grain boundary appears to be quite well described by the Gaussian function, although the buckling is wider at the base. In contrast, the buckling of the $17.9^\circ$ boundary is both wider at the base and noticeably narrower at the peak compared to the Gaussian with FWHM = \unit[5]{nm}. The narrower Gaussian with FWHM = \unit[2]{nm}, while completely failing to reproduce the shape at the base, describes the behavior near the peak of the buckling fairly well.

 Considering the greater success of the continuum model with FWHM = \unit[2]{nm} compared to that with FWHM = \unit[5]{nm}, the shape of the buckling near the peak appears to be important for the scattering. This can be explained using Equations (\ref{eq:wavew_leading}) and (\ref{eq:waveu_leading}). In these equations it is seen that the scattering is due mainly to the second derivative of $w_0(x)$, i.e. the curvature of the buckling. The largest curvature of the actual grain boundary buckling occurs at the peak, and this curvature will effectively determine the scattering strength. To obtain a good correspondence between MD and the continuum mechanical model it is therefore important to have a scattering center that reproduces the peak curvature given by the atomistic simulations. In the case of the $17.9^\circ$ boundary this is achieved with FWHM = \unit[2]{nm}, but not with FWHM = \unit[5]{nm}. 

\section{Conclusion}
In the present study, molecular dynamics simulations have been used to investigate the scattering of long-wavelength flexural (out-of-plane) phonons against grain boundaries in graphene. Of the three considered grain boundaries, the one with a misorientation angle of $32.2^\circ$ is flat and the other two, with misorientation angles $9.4^\circ$ and $17.9^\circ$, display a substantial out-of-plane buckling. The buckling of the $9.4^\circ$ boundary is found to be \unit[0.6]{nm} high and \unit[1.7]{nm} wide, while that of the $17.9^\circ$ boundary is \unit[1.5]{nm} high and \unit[5]{nm} wide. Due to the slow convergence of the buckling characteristics with respect to system size, previous studies\cite{Carlsson2011gbstruct,Liu2011gbstruct} on smaller systems have found lower buckling heights.

The results of the phonon scattering simulations show large differences between flat and buckled boundaries. At the flat $32.2^\circ$ boundary, the transmission is over \unit[95]{\%} for wavelengths above \unit[1]{nm}, indicating that this boundary does not significantly scatter long-wavelength flexural phonons. In contrast, both buckled boundaries are seen to cause substantial scattering: for the $17.9^\circ$ boundary the transmission reaches a minimum value of \unit[20]{\%} at wavevector $k_0=1$~nm$^{-1}$, while for the $9.4^\circ$ boundary the minimum transmission is \unit[40]{\%} at $k_0=1.5$ nm$^{-1}$. Additionally, the buckled boundaries scatter between $20$ and \unit[50]{\%} of the total vibrational energy into the longitudinal mode. Clearly, it is the buckling of the grain boundaries that scatters long-wavelength flexural phonons. 

It has been claimed that long-wavelength flexural phonons contribute significantly to the thermal conductivity of graphene \cite{Seol2010scienceZA,Lindsay2010thermal}. If this is the case, the phonon scattering results above indicate that the boundary buckling should have a substantial influence on the thermal conductivity across grain boundaries, the Kapitza conductance. Such an effect seems especially likely given that buckling due to compressive strain has been found to change the thermal conductivity of graphene\cite{Wei2011strain,Li2010strain}. Unfortunately, the existing studies of the Kapitza conductance in graphene do not mention grain boundary buckling at all\cite{Cao2012Kapitza,Cao2012asymmetric,Bagri2011,Lu2012Greens,Serov2013greens}, although some consider the effects of buckling due to strain \cite{Liu2013strain,Tang2013strain}. It is possible that no significant buckling has been seen in these studies since they frequently use a fixed system size, something that may reduce or eliminate buckling by preventing the system from contracting in the direction perpendicular to the boundary. An investigation of the dependence of the Kapitza conductance on boundary buckling could supply new insights and contribute to the understanding of the thermal conductivity in grain boundaries subject to compressive strain. 

In addition to the molecular dynamics simulations a continuum mechanical model of the phonon scattering has been constructed. The system is modeled as a string restricted to vibrating in one transversal mode and the longitudinal mode, with the stress effectively determined by a static transversal displacement representing the buckling. The continuum mechanical model shows good qualitative agreement with the molecular dynamics results. For the $9.4^\circ$ boundary the transmission minimum at $k_0=1.5$ nm$^{-1}$ is reproduced with surprising accuracy when the static transversal displacement is given the same height and width as the observed boundary buckling. To reach a similar agreement for the $17.9^\circ$ grain boundary it is necessary to reduce the width of the static displacement by \unit[3]{nm} compared to the width of the boundary buckling, which gives a better description of the buckling curvature at the peak. The results show that continuum mechanics can be used to describe the effect of buckled grain boundaries on long-wavelength flexural phonons in graphene.

\newpage
\section*{Acknowledgements}
We would like to thank Prof. Jari Kinaret for stimulating discussions. We also acknowledge financial support from the Swedish Research Council (VR) and the EU Graphene Flagship (grant no. 604391).

\end{document}